\documentclass[10pt,
               aps,
               pre,
               twocolumn,
               showpacs, 
               groupedaddress]{revtex4-2}
\usepackage[utf8]{inputenc}
\usepackage{graphicx}

\usepackage[dvipsnames,table]{xcolor}
\usepackage[caption=false]{subfig}
\usepackage{amsmath, amsfonts, amssymb}
\usepackage[linktocpage=true,
  colorlinks=true, 
  pdfborder={0 0 0},
  linkcolor=blue,
  citecolor=blue,
  filecolor=yellow,
  urlcolor=blue,
  bookmarks,
  pdfauthor={},
]{hyperref}
\usepackage{tikz}
\graphicspath{{images/}}

\makeatletter
\newcommand*{\rom}[1]{\expandafter\romannumeral #1}
\makeatother

\newcommand{\Teff}{\ensuremath{T_{\text{eff}}}}

\begin{document}

\title{Self-propulsion and self-rotation of an inertial chiral active Ornstein-Uhlenbeck particle}
\author{F Sahala}
\affiliation{Department of Physics, University of Kerala, Kariavattom, Thiruvananthapuram-$695581$, India}

\author{M Muhsin}
\affiliation{Department of Physics, University of Kerala, Kariavattom, Thiruvananthapuram-$695581$, India}

\author{M Sahoo}
\email{jolly.iopb@gmail.com}
\affiliation{Department of Physics, University of Kerala, Kariavattom, Thiruvananthapuram-$695581$, India}

\date{\today}
\begin{abstract}
We investigate the transport feature of an inertial chiral active Ornstein-Uhlenbeck particle moving on a two-dimensional surface. Using both analytical approach and numerical simulations, we have exactly explored the transient and steady-state behavior of the particle by analyzing the simulated particle trajectories, probability distribution functions for position and velocity, mean square displacement, mean square velocity, and effective kinetic temperature of the medium. From the mean square displacement calculations, we observe that, unlike an inertial active Brownian particle, a chiral active particle manifests an initial ballistic, intermediate sub-diffusive to non-diffusive, and the conventional long-time diffusive behavior. The intermediate sub-diffusive to non-diffusive behavior is prominent for the self-propulsion of an overdamped particle. It can be understood by chirality-induced transient confinement, which persists for short time intervals and diffuses away in the time asymptotic limit or at the steady state. This behavior is further complemented by the exact calculation of mean square velocity or effective kinetic temperature of the medium, which is a decreasing function of the magnitude of chirality. Moreover, the steady-state MSD and MSV are found to have a dependence both on chirality and activity time scale and hence can be controlled by tuning the persistent duration of activity or strength of the chirality of the particle.
\end{abstract}

\maketitle
\section{INTRODUCTION}\label{sec:intro}
In recent years, research on active matter has become a rapidly emerging area that spans multiple disciplines including physics, biology, chemistry, and engineering~\cite{marchetti2013hydro, Ramaswamy2017active, Gompper2020roadmap}. 
Unlike passive systems for which external bias is needed for the propulsion of particles in the medium or environment, these self-propelling active systems can move and organize on their own, often leading to a complex and collective behavior~\cite{bechinger2016active}. 
The field of active matter spans a wide range of scales and systems, from biological entities like flocks of birds~\cite{mora2016local}, bacterial colonies~\cite{peruani2012collective}, and motor proteins~\cite{Gompper2020roadmap} to artificial synthetic systems like self-propelled colloids and active gels~\cite{walther2013janus, howse2007self}.

The breaking of translational symmetry induces self-propelled motion in the constituents of active matter, resulting in persistent motion along stochastic directions.
While many active systems primarily exhibit linear self-propulsion, certain natural and artificial systems also demonstrate circular motion due to intrinsic particle asymmetry. 
This circular motion is attributed to the breaking of orientational symmetry that classifies these systems as chiral active matter~\cite{liebchen2022chiral}. 
The components of chiral active systems possess intrinsic chirality, enabling them both to self-propel and self-rotate.
Examples of chiral active matter include biological entities such as molecular motors~\cite{loose2014bacterial, afroze2021monopolar}, swimming bacteria~\cite{friedrich2008stochastic, diluzio2005ecoli, lauga2005swimming, leonardo2011swimming}, sperm cells~\cite{riedel2005self, friedrich2007chemotaxis}, and so on. 
In addition to these, artificial systems such as colloidal microswimmers with shape asymmetry~\cite{kummel2013circular, ten2014gravitaxis, wang2017janus, shelke2019transition, zhang2020reconfigurable} and rotating nano and micro-robots~\cite{barois2020sorting, nourhani2013chiral} also possess chirality.
Chirality plays a critical role in shaping the behavior of such systems at both the individual and collective levels, significantly influencing the dynamics, stability, and their phase behavior~\cite{kreienkamp2022clustering, keaveny2009hydro, liao2018clustering, vivek2024macro, levis2018micro, lowen2016chirality, caprini2024self}.

Modeling chiral active particles (CAPs) can be achieved by incorporating chirality into well-established frameworks for self-propelled particles. For instance, in models like Active Brownian Particles (ABPs)~\cite{hagen2011brownian, cates2013when, kanaya2020steady, stenhammar2014phase, solon2015pressure, caprini2021collective, caprini2020hidden, scholz2018inertial}
or Active Ornstein-Uhlenbeck Particles (AOUPs)~\cite{lehle2018analyzing, bonilla2019active, martin2021statistical, caprini2019active, caprini2021inertial, dabelow2019irreversibility, berthier2019glassy, wittman2018effective, fily2019self, mandal2017entropy, fodor2016far, muhsin2021orbital, muhsin2022inertial, adersh2024transition}, the chirality can be represented by coupling translational and rotational degrees of freedom or by introducing torque-like effects~\cite{caprini2023chiral}. Such models provide a flexible platform to explore the rich dynamics of chiral active matter under various conditions, including confinement, external fields, and interactions with other particles or boundaries.
The behavior of CAPs near boundaries and in confined geometries reveals intriguing phenomena such as surface currents and bulk accumulation~\cite{caprini2019activechiral} and the presence of $x$-$y$ correlation in non-radial potentials~\cite{caprini2023chiral}. 
Furthermore, chirality has been found to suppress non-Gaussianity in confined Active Brownian Particles~\cite{caprini2023chiral}. 
Despite these advances, the majority of research on CAPs has been mostly concentrated on overdamped systems~\cite{sevilla2016diffusion, liebchen2022chiral, mite2013sorting, chen2018unidirectional, meng2020transport}, while the inertial dynamics remain largely unexplored.

In the present work, we investigate the dynamics of an inertial chiral active Ornstein-Uhlenbeck particle (AOUP) employing both analytical techniques and numerical simulations. Using both analytical approaches and numerical simulations, we explore the transient and steady-state behavior of the particle by analyzing the particle trajectories, probability distribution functions for position and velocity, mean square displacements (MSD), mean square velocity (MSV), and the effective kinetic temperature of the medium. From the MSD calculations, it is interestingly observed that in contrast to normal active particle, the MSD of a chiral particle displays an intermediate subdifussive to nondifussive behaviour with increase in magnitude of chirality. This intermediate time non-difussive regime is more pronounced for the case of an overdamped particle, where the inertia has negligible impact. The intermediate time non-difussive feature of the particle clearly predicts the existence of a chirality induced transient confinement, which disappears in the time-asymptotic limit or at steady state. This feature is further supported by the exact calculation of mean square velocity and effective kinetic temperature of the medium, which decrease as a function of magnitude of chirality.
In the next section, we discuss the model under consideration and introduce the physical quantities of interest. It is then followed by a detailed discussion of results, and finally we conclude.

\section{MODEL AND METHOD}\label{sec:model}
We consider the motion of an inertial chiral active Ornstein-Uhlenbeck particle of mass $m$, self-propelling in a two-dimensional (2d) plane. The dynamics of the particle can be described by the Langevin equation of motion~\cite{muhsin2021orbital,caprini2019activechiral}
\begin{equation}
m \ddot{\bf{r}} (t) = - \gamma \dot{\bf{r}} (t) + \boldsymbol{{\xi}} (t) + \sqrt{2 \gamma k_BT} \boldsymbol{\eta}(t).
\label{eq:model dynamics}
\end{equation}
Here, ${\bf r}(t) = x(t) \hat{i} + y(t) \hat{j}$ represents the position vector and $\gamma$ is the viscous coefficient of the medium. 
The noise $\boldsymbol{\eta}(t)$ is a zero averaged, delta-correlated white noise. It models the thermal fluctuations of the bath kept at temperature $T$. The stochastic term $\boldsymbol{\xi}(t)$ represents the active force responsible for the self-propulsion of the particle. Here, we model this force as a chiral extension of the Ornstein-Uhlenbeck process~\cite{caprini2019activechiral}. Hence, the dynamics of $\boldsymbol{\xi}(t)$ is given by
\begin{equation}
    \boldsymbol{\dot{\xi}}(t) = - \frac{1}{t_c} \boldsymbol{\xi}(t) + \xi_0 \sqrt{\frac{2}{t_c}}  \boldsymbol{\zeta}(t) +  \omega\boldsymbol{\xi}(t) \times\bf\hat{k}.
    \label{eq:model dynamics2}
\end{equation}
\begin{figure}[!ht]
    \centering
  \includegraphics[width=\linewidth]{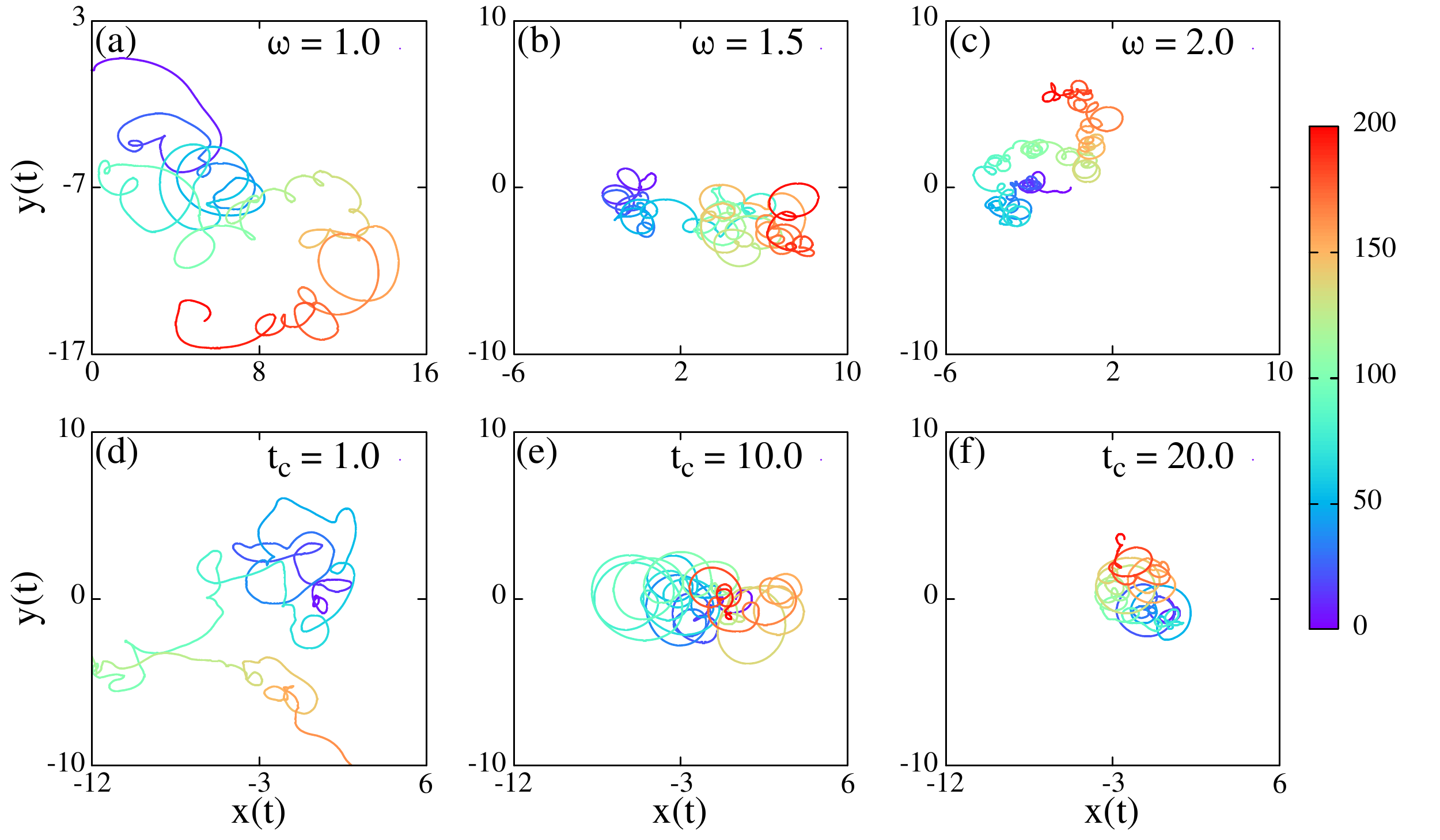}
    \caption{The simulated particle trajectories in the $x-y$ plane are plotted in (a),(b), and (c) for different values of $\omega$, keeping fixed $t_c=5$ and in  (d), (e), and (f) for different values of $t_c$, keeping fixed $\omega=1$. The other common parameters are $t_m =  \gamma = \xi_0 = 1$ and $T = 0.01$. The color map shows the time evolution of the trajectories.}
    \label{fig:Traj_omag_tc}
\end{figure}
 Here, $t_c$ represents the self-propulsion or activity timescale of the system, $\xi_0$ is the strength of self-propulsion force, and $\boldsymbol{\zeta}(t)$ is a delta-correlated white noise with zero average and unit variance. 
 The chirality is introduced into the dynamics via the term $\omega\boldsymbol{\xi}(t) \times\bf\hat{k}$~\cite{caprini2023chiral, caprini2019activechiral}, with the parameter $\omega$ quantifying the magnitude of chirality of the particle. 
 The unit vector $\hat{k}$ is along the positive direction of $z$-axis. This term is always directed perpendicular to $\boldsymbol{\xi}(t)$ in the $x-y$ plane. Hence, the effect of this term is to change the particle's orientation. The active force $\boldsymbol{\xi }(t)$ in Eq.~\eqref{eq:model dynamics2} has the statistical properties
 \begin{align}
 \langle \xi_i(t) \rangle &= 0 , \quad \text{and}\nonumber\\ 
 \langle \xi_i(t) \xi_j(t') \rangle &= \xi_0^2  \delta_{ij} \exp\left(-\frac{|t - t'|}{t_c}\right) \cos{\omega (t-t')}. 
 \label{eq:xi(t)}
 \end{align}

Two limiting cases of Eq.~\eqref{eq:model dynamics2} are noteworthy. In either $t_c \to 0$ limit or $\omega \to \infty$ limit, one can see that Eq.~\eqref{eq:model dynamics2} becomes $\boldsymbol{\xi} = 0$, and hence the dynamics Eq.~\eqref{eq:model dynamics} reduces to the case of a passive Brownian particle. The mass $m$ and Boltzmann constant $k_B$ are considered as unity throughout the rest of this paper. Next, we define the physical quantities of interest. The MSD of the particle is given by  
\begin{equation}
    \langle \Delta{\bf r}^2(t) \rangle = \langle ({\bf r}(t) - {\bf r}(0))^2 \rangle.
    \label{eq:msd-def}
\end{equation}
Similarly, MSV can be obtained as
\begin{equation}
    \langle \Delta{\bf v}^2(t) \rangle = \langle ({\bf v}(t) - {\bf v}(0))^2 \rangle.
    \label{eq:msv-def}
\end{equation}

We have also simulated the dynamics Eq.~\eqref{eq:model dynamics} using a second-order Euler scheme. The simulation is run up to $10^3$ time steps and the duration of each time step is taken to be $10^{-3}$. The steady-state measurements are taken averaging over $10^5$ realizations and after neglecting the initial transients of $10^2$ time steps for each realization.

\section{RESULTS AND DISCUSSION}\label{sec:result}
Solving 
Eq.~\eqref{eq:model dynamics} with initial conditions
${\bf r}(0) = {\bf r_0} $ and ${\bf \dot{r}}(0) = {\bf v_0}$, we obtain the position ${\bf r}(t)$ as
\begin{equation}
    {\bf{r}}(t) = {\bf{r_0 }}+ {\bf{v_0}} t_m\left(1-e^{-\frac{t}{t_m}}\right) + \int\limits_{0}^{t} \frac{1-e^{-\frac{t-t'}{t_m}}}{\gamma}{\boldsymbol{\widetilde{\xi}}}(t') dt',
    \label{eq:r(t)}
\end{equation}
and the velocity ${\bf v}(t)$ as
\begin{align}
    {\bf{v}}(t) =& {\bf{v_0 }} e^{-\frac{t}{t_m}} +  \frac{1}{m}\int\limits_{0}^{t} e^{-\frac{t - t'}{t_m}} \boldsymbol{\widetilde{\xi}}(t')\;dt',
    \label{eq:v(t)}
\end{align}
with $\boldsymbol{\widetilde{\xi}}(t) = \boldsymbol{\xi}(t) + \sqrt{2\gamma k_BT}\boldsymbol{\eta}(t)$. Here $t_m = \frac{m}{\gamma}$ is the inertial time scale.
In Fig.~\ref{fig:Traj_omag_tc}, 
we have shown the simulated trajectories [Eq.~\eqref{eq:model dynamics}] of a chiral active particle in $x-y$-plane for different values of $\omega$ [Figs.~\ref{fig:Traj_omag_tc}(a), (b) and (c)] and $t_c$ [Figs.~\ref{fig:Traj_omag_tc}(d), (e) and (f)]. The trajectories show a self-rotation along with the self-propulsion for finite values of $t_c$ and $\omega$. 
The chirality $\omega$ can be regarded as the frequency of these circular motions, and hence related to the self-propulsion velocity of the particle $v_s$ as $\omega \approx \frac{v_s}{R}$, where $R$ is the radius of the circular motion~\cite{caprini2023chiral}. Thus, with an increase in chirality, the radius of circular trajectories decreases. This can be confirmed from Figs.~\ref{fig:Traj_omag_tc}(a), (b), and (c), where an increase in the value of $\omega$ causes smaller circles to appear in the trajectory.
Further, the trajectory of the particle gets confined to a smaller region in the $x-y$ plane. 
However, with an increase in the activity timescale $t_c$, the number of circles in the trajectory increases, and the motion of the particle is restricted to a small region [Figs.~\ref{fig:Traj_omag_tc}(d), (e) and (f)].
These observations suggest a transient confinement of the particle with chirality $\omega$ and activity time scale $t_c$.

\begin{figure}[!hb]
    \centering
  \includegraphics[width=\linewidth]{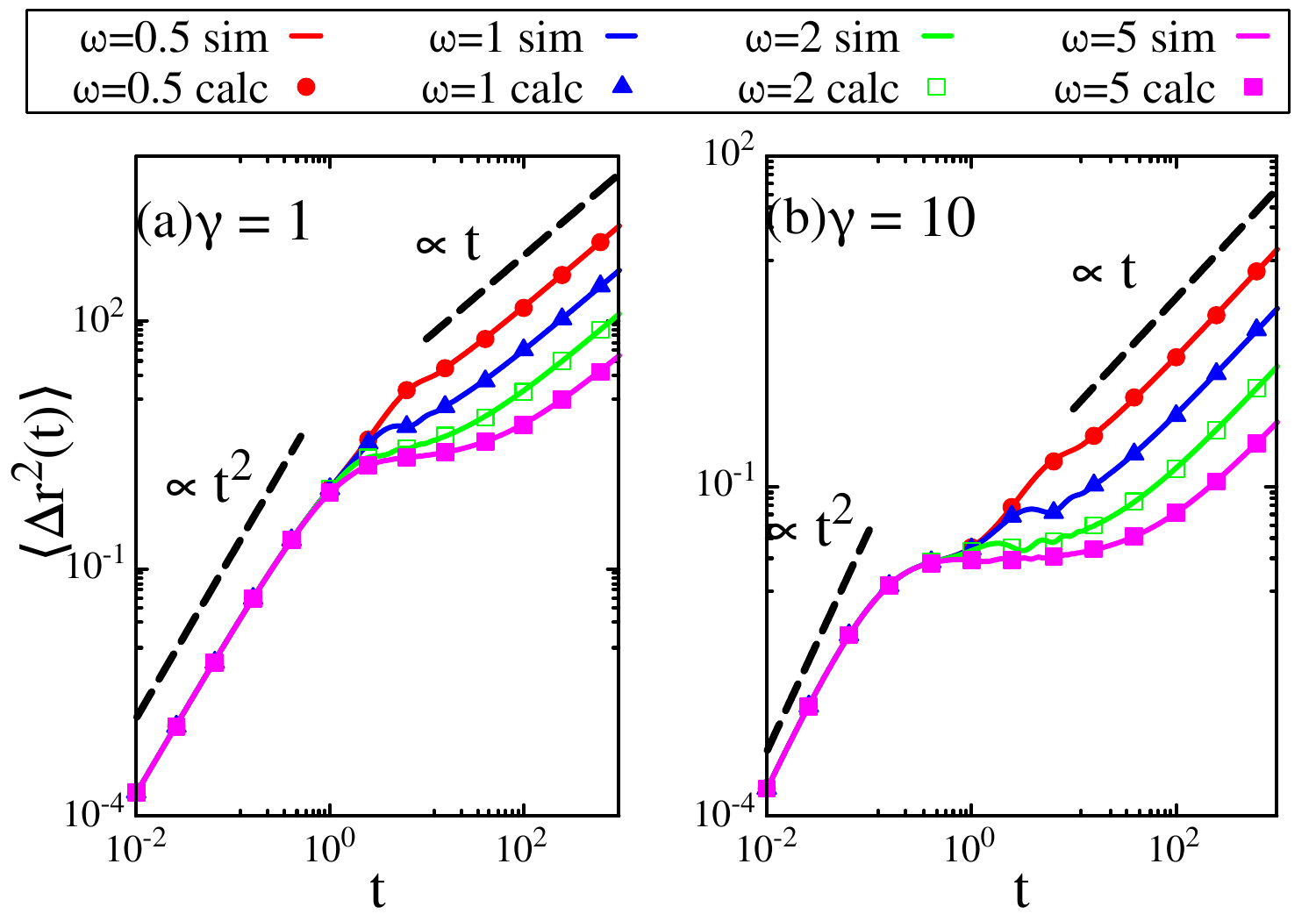}
    \caption{The plot of MSD [Eq.~\eqref{eq:msd1}] as a function of $t$ for different values of $\omega$ is shown in (a) for $\gamma = 1.0$, and in (b) for $\gamma = 10.0$. The solid lines correspond to the calculation, and the symbols represent the simulated results. The other common parameters are $t_m =\xi_0  =1, t_c =5$, and $T =0.01$.}
    \label{fig:MSD_omega}
\end{figure}

Next, we have exactly calculated the MSD using Eq.~\eqref{eq:msd-def}. Substituting the solution of ${\bf r}(t)$ from Eq.~\eqref{eq:r(t)} in Eq.~\eqref{eq:msd-def}, we obtain
\begin{widetext}
\begin{equation}
\begin{split}
  \langle \Delta{\bf r}^2(t) \rangle &= v_0^2 \left(e^{-\frac{t}{t_m}}-1\right)^2 t_m^2 + \frac{2k_BT}{\gamma }\left[\left(-e^{-\frac{2 t}{t_m}}+4 e^{-\frac{t}{t_m}}-3\right) t_m+2 t\right] +  \frac{2\xi _0^2 t_c}{\gamma ^2} \Bigg\{
  \frac{2(t + t_c - t_m)}{\Omega} - \frac{4 t_c}{\Omega^2} - \frac{t_c + t_m}{t_c^2|\nu_{+}|^2}\\
  & - \frac{2(t_c - 2 t_m)e^{-\frac{t}{m}}}{\Omega|\nu_{-}|^2}\left( \frac{1}{t_c^2} - \frac{1}{t_c t_m} + \omega^2 \right) + \frac{2e^{-\frac{t}{t_c}}}{t_m^2 |\nu_{-}|^2 \Omega^2}\left[ t_c - t_m - t_c^2(t_c - 3t_m)\omega^2  \right]\cos\omega t + \frac{e^{-\frac{2t}{t_m}}(t_c - t_m)}{t_c^2 |\nu_{-}|^2} \\
  & - \frac{2e^{-\frac{t}{t_c}}}{t_m^2 |\nu_{-}|^2 \Omega^2}\left[ -3t_m + t_c\left( 2 + t_c t_m\omega^2 \right) \right]t_c \omega\sin\omega t - \frac{2e^{-t \left( \frac{1}{t_c} + \frac{1}{t_m} \right)}}{t_m |\nu_{+}|^2|\nu_{-}|^2\Omega}\left[ 2 \omega^2 + \Re\left( \nu_{+} \nu_{-}^* \right)\right]\cos\omega t \\
  & + \frac{2e^{-t \left( \frac{1}{t_c} + \frac{1}{t_m} \right)}}{t_c t_m |\nu_{+}|^2|\nu_{-}|^2\Omega}\left[-2\omega + t_c^2\omega\Re\left( \nu_{+} \nu_{-}^* \right) \right]\sin\omega t
  \Bigg\},
\end{split}
\label{eq:msd1}
\end{equation}
\end{widetext}
with $\Omega = 1 + t_c^2 \omega^2$, $\nu_{+} = \omega + i \left(\frac{1}{t_m} + \frac{1}{t_c}\right)$, and $\nu_{-} = \omega + i \left(\frac{1}{t_m} - \frac{1}{t_c}\right)$. Here, $\Re(\cdots)$ represents the real part and `*' represents the complex conjugate.
In Fig.~\ref{fig:MSD_omega}, we have plotted $\langle \Delta {\bf r}^2(t) \rangle$ as a function of time for different values of $\omega$ for $\gamma=1$ in Fig.~\ref{fig:MSD_omega}(a) and for $\gamma=10$ in Fig.~\ref{fig:MSD_omega}(b), respectively.
The MSD exhibits an initial time ballistic regime (i.e., $\langle \Delta {\bf r}^2(t) \rangle \propto t^2$), and a diffusive steady state (i.e., $\langle \Delta {\bf r}^2(t) \rangle \propto t$). The expression of MSD for lower time regime is given by $\langle \Delta{\bf r}^2(t) \rangle = v_0^2t^2 + O(t^3)$, which shows the ballistic regime as discussed. Also, MSD in this regime does not have a dependence on $t_c$ or $\omega_c$, as shown in Fig.~\ref{fig:MSD_omega}.
In the steady-state, the value of MSD is diffusive and is given by
\begin{align}
  \langle \Delta{\bf r^2 \rangle_s} &= \lim_{t\to \infty}\langle \Delta{\bf r^2 \rangle}\nonumber\\ 
  &= \frac{4 t}{\gamma ^2}\left(\gamma  k_BT + \frac{\xi _0^2 t_c}{\omega ^2 t_c^2+1}\right).
  \label{eq:MSD-steady}
\end{align}
It is clear from Eq.~\eqref{eq:MSD-steady} that the steady state MSD depends both on $\omega$ and $t_c$. In the $t_c \to 0$ limit, MSD recovers the equilibrium value [$\langle \Delta{\bf r^2 \rangle_s}=  \frac{4 k_B T t}{\gamma}$]. This is because, the limit of vanishing activity time scale corresponds to a passive system in equilibrium, and hence the fluctuation-dissipation relation is validated. The same result is recovered for a very high value of chirality (i.e., in the limit $\omega \to \infty$). 
Furthermore, as the value of $\omega$ increases, the steady-state MSD decreases, complementing with the suppression of particle trajectories that are observed in Fig.~\ref{fig:Traj_omag_tc}.
Interestingly, the MSD displays an intermediate-time unusual plateau behavior as the value of $\omega$ is increased [Fig.~\ref{fig:MSD_omega}(a)]. In this regime, the MSD goes into a temporal kinetic arrest, which suppresses the overall displacement of the particle. This plateau behavior is more pronounced for $\gamma=10$, i.e., in the overdamped regime [Fig.~\ref{fig:MSD_omega}(b)], where the inertial influence is negligible.

In order to investigate this intermediate time behavior of MSD, we define the MSD exponent $\alpha$ such that $\langle \Delta {\bf r}^2(t) \rangle \propto t^\alpha$. That is,
\begin{equation}
    \alpha = \frac{\partial \log \langle \Delta {\bf r}^2(t) \rangle}{\partial \log t}.
    \label{eq.msd_alpha}
\end{equation}
\begin{figure}
    \centering
  \includegraphics[width=0.8\linewidth]{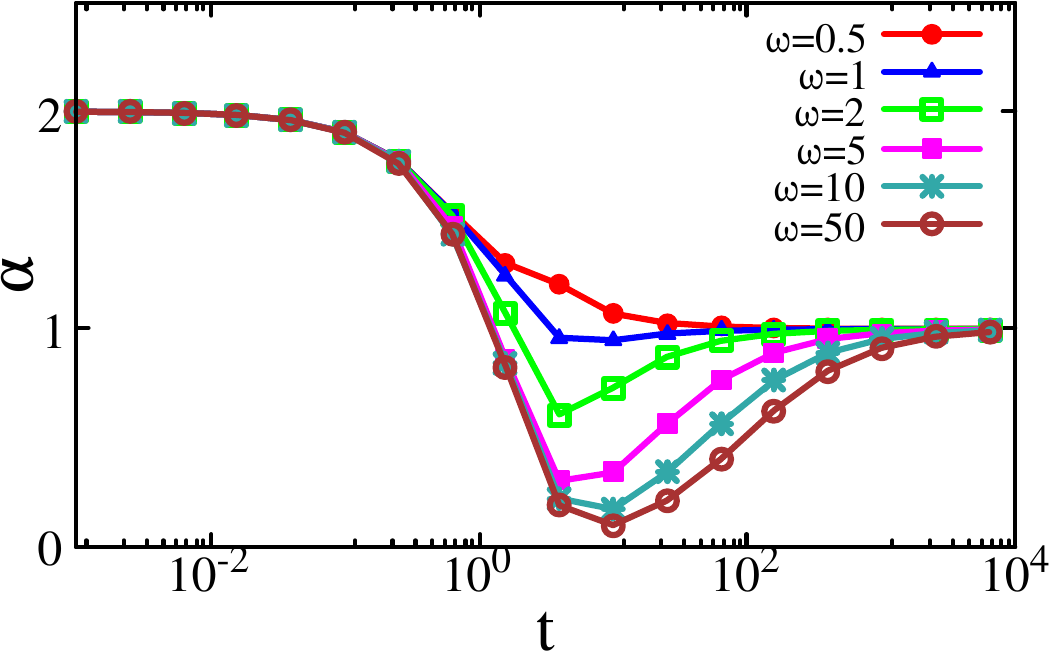}
    \caption{The plot of MSD Exponent $\alpha$ [Eq.~\eqref{eq.msd_alpha}] as a function of $t$ for different values of  $\omega$. The other common parameters are $t_m =\xi_0 =\gamma =1$, $t_c = 5$, and $T =0.01$.}
    \label{fig:MSD_alpha}
\end{figure}
\begin{figure}
    \centering   \includegraphics[width=\linewidth]{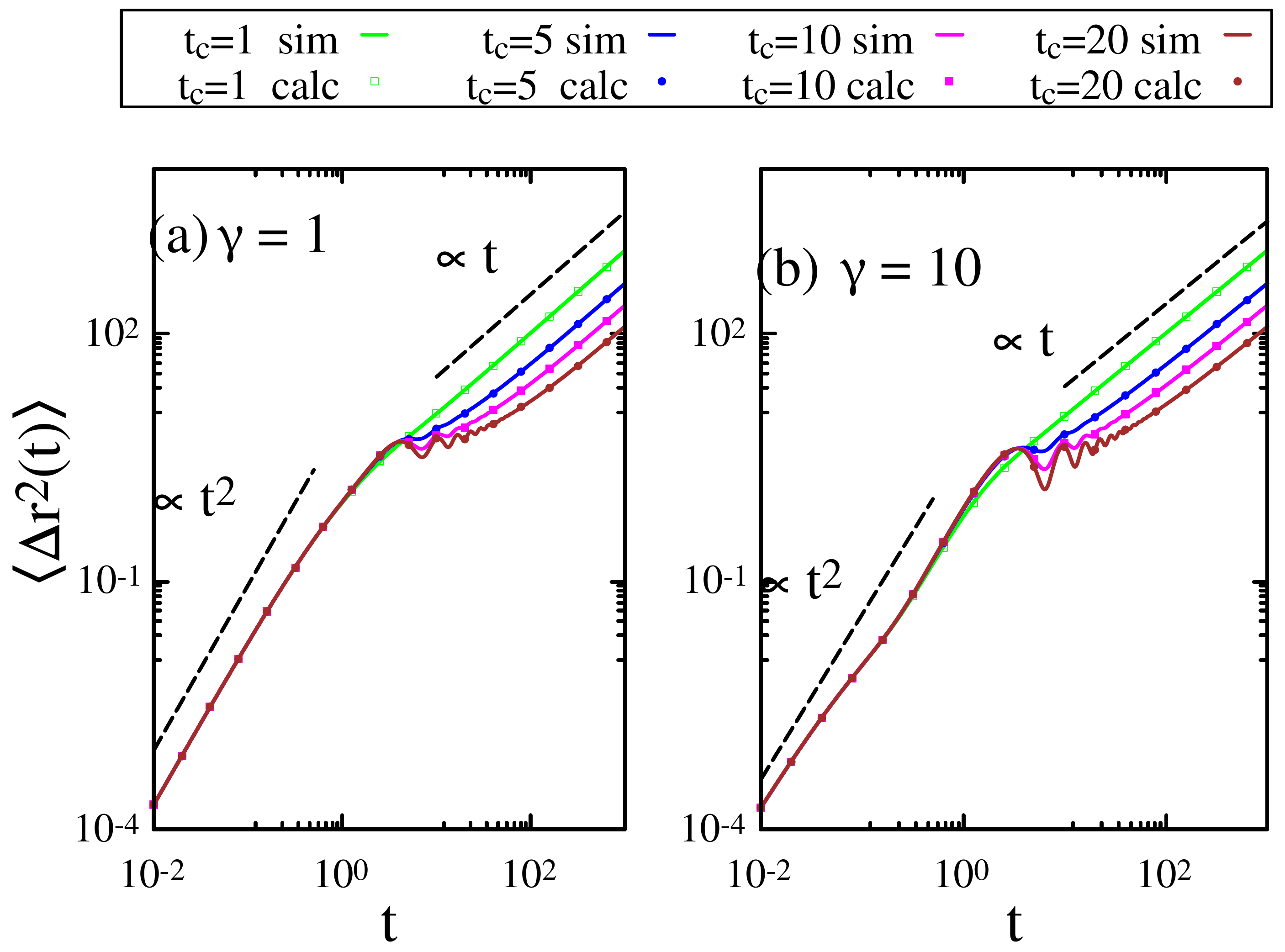}
    \caption{The plot of MSD from Eq.~\eqref{eq:msd1} and from simulation  as a function of $t$ for different values of $t_c$ is shown in (a) for $\gamma = 1.0$ and in (b) for $\gamma = 10.0$. The other common parameters are $t_m =\xi_0 =\omega  =1$ and $T =0.01$. }
    \label{fig:MSD_tc}
\end{figure}
Thus, $\alpha$ represent the instantaneous slope of $\log \langle \Delta {\bf r}^2(t) \rangle$ versus $\log t$ curve. 
The variation of $\alpha$ with time corresponding to Fig.~\ref{fig:MSD_omega}(a) is shown in Fig.~\ref{fig:MSD_alpha}. 
In the short time regime (i.e., $t\rightarrow 0$ limit), the exponent $\alpha = 2$ indicates the ballistic behavior of MSD. In the steady state, the MSD is diffusive with $\alpha = 1$, as discussed. 
In the intermediate time regime, for small chirality ($\omega = 0.5$ in Fig.~\ref{fig:MSD_alpha}), $\alpha$ monotonically decay from ballistic to diffusive behavior. However, with an increase in the value of $\omega$, the intermediate time behavior of MSD changes from sub-diffusive ($\alpha<1$) to non-diffusive ($\alpha= 0$) before approaching the diffusive behavior with $\alpha=1$ in the steady state. 
This behavior of $\alpha$ indicates a transient confinement induced by chirality, confirming the temporal kinetic arrest of MSD~\cite{berthier2011theoratical, ai2024rotational}.
Further, for a very high value of chirality, that is, as $\omega\rightarrow \infty$, the value of $\langle \Delta{\bf r}^2 \rangle_{s}$ becomes $\frac{4 k_BT t}{\gamma }$, which is independent of $t_c$ and is same as the steady state value of MSD for a passive Brownian particle. 

In Fig.~\ref{fig:MSD_tc}, we have plotted $\langle \Delta{\bf r}^2(t) \rangle$ as a function of time for different values of $t_c$ for $\gamma=1$ in Fig.~\ref{fig:MSD_tc}(a) and for $\gamma=10$ in Fig.~\ref{fig:MSD_tc}(b), respectively. The MSD does not show any dependence on activity in the initial time ballistic regime. In the long-time diffusive regime, the MSD decreases with increase in $t_c$ value, accounting for the suppression of particle trajectories due to the appearance of rotational motion. 
In the limit  $t_c \to \infty$, the value of $\langle \Delta{\bf r}^2 \rangle_{s}$ is $\frac{4 k_BT t}{\gamma}$, which is the steady state MSD for a passive Brownian particle. In the intermediate time regime, oscillations start to appear with increase in the value of $t_c$, as observed in Ref.~\onlinecite{pattanayak2024impact}. The amplitude of such oscillations is significant in the overdamped case [Fig.~\ref{fig:MSD_tc}(b)]. The emergence of such oscillations in the intermediate time regime might be attributed due to the complex interplay between the chirality and activity.
\begin{figure*}
    \includegraphics[width=0.8\linewidth]{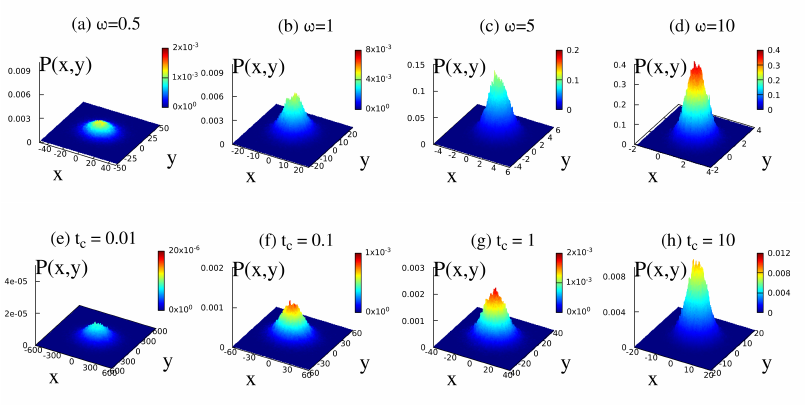}
    \caption{The simulated probability distribution  of position at time $t = 100$ for different values of $\omega$ fixing $t_c=5.0$ are shown in (a)-(d) and for different values of $t_c$ fixing $\omega=1.0$ are shown in (e)-(h). Other parameters are $t_m = \gamma = \xi_0 = 1$ and $T = 0.01$.}
    \label{fig:PD_x}
\end{figure*}

In Fig.~\ref{fig:PD_x}, we have plotted the simulated position probability distribution for different values of $\omega$ [Fig.~\ref{fig:PD_x} (a), (b), (c), and (d)] and $t_c$ [Fig.~\ref{fig:PD_x}(e), (f), (g) and (h)]. The distribution is Gaussian and is centered at the origin, due to the linearity of the model and the Gaussian nature of the active force. As $\omega$ and $t_c$ increase, the width and hence the variance of the distribution decreases and narrows out [see Fig.~\ref{fig:PD_x}]. At the same time, the peak of the distribution is enhanced. This implies that with an increase in magnitude of chirality and activity timescale, the probability of finding the particle at the mean position increases and that of finding the particle at far away distances from the center decreases, supporting the results that obtained from the particle trajectories and MSD calculations.  

\begin{figure}
    \centering
\includegraphics[width=\linewidth]{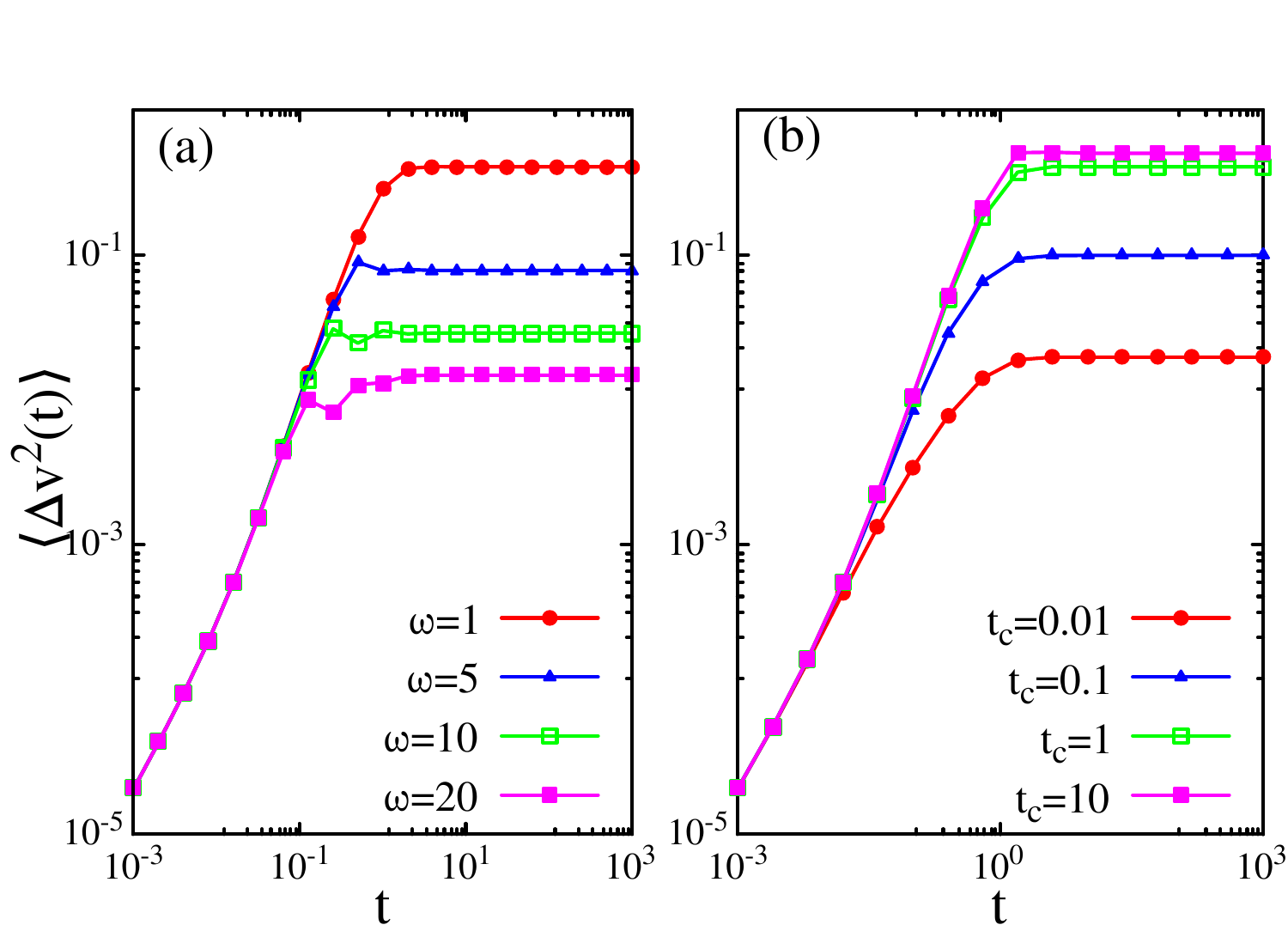}
    \caption{The plot of MSV from Eq.~\eqref{eq:msv1}  as a function of $t$  for different values of $\omega$ fixing $t_c=5.0$ is shown in (a) and for different values of $t_c$ for a fixed $\omega=1.0$ is shown in (b). The other common parameters are $t_m =\gamma = \xi_0 =1$ and $T = 0.01$. }.
    \label{fig:MSV_tc_omega}
\end{figure}

Next, we have exactly  calculated the mean square velocity (MSV) [Eq.~\eqref{eq:msv-def}] of the particle, which is given by
\begin{widetext}
\begin{equation}
\begin{split}
\langle \Delta{\bf v}^2(t) \rangle =& v_0^2 \left(e^{-\frac{2 t}{t_m}} - 1\right)^2 + \frac{4 \xi _0^2} {m^2 t_c} \frac{e^{-t\left( \frac{1}{t_c} + \frac{1}{t_m} \right)}}{|\nu_{+}|^2|\nu_{-}|^2} \Bigg\{ -t_c\cos(\omega t)\Re \left(\nu_{+}\nu_{-}^* \right) - 2\omega\sin(\omega t) \\
& + e^{\frac{t}{t_c}}\left[ t_c \Re \left(\nu_{+}\nu_{-}^* \right)\cosh\left( \frac{t}{t_m}\right) + t_m\Re \left(\nu_{+}\nu_{-} \right) \sinh \left( \frac{t}{t_m} \right) \right] \Bigg\} + \frac{2k_BT \left(1-e^{-\frac{2 t}{t_m}}\right)}{\gamma  t_m}.
\label{eq:msv1}
\end{split}
\end{equation}
\end{widetext}
We have plotted $\langle \Delta{\bf v}^2(t) \rangle$ as function of time [Eq.~\eqref{eq:msv1}] for different values of $\omega$ in Fig.~\ref{fig:MSV_tc_omega}(a) and for different values of $t_c$ in Fig.~\ref{fig:MSV_tc_omega}(b). 
In the initial-time regime, the MSV is independent of both $\omega$ and $t_c$ values. The steady-state MSV is suppressed with an increase in the value of $\omega$, while it is enhanced with an increase in the value of $t_c$.
This steady state MSV ($\langle \Delta{\bf v}^2 \rangle_s $) can be obtained from Eq.~\eqref{eq:msv1} by taking $t\rightarrow \infty$ limit, that is in the steady state. It is given by
\begin{equation}
\begin{split}
  \langle \Delta{\bf v}^2 \rangle_s &= \lim_{t \to \infty} \langle \Delta{\bf v}^2 (t)\rangle \\
  &= \frac{2} {\gamma ^2}\left(\frac{\xi _0^2 t_c \left(t_c+t_m\right)}{t_c^2 \left(\omega ^2 t_m^2+1\right)+2 t_c t_m+t_m^2}+\frac{\gamma  k_BT}{t_m}\right).
\end{split}
    \label{eq:msv_steady_limit}
\end{equation}
The steady-state MSV has two parts. The first part is due to the contribution from the self-propulsion of the particle and the second part is exactly the same as that of a passive Brownian particle. In either of the limit $t_c \to 0$ or $\omega \to \infty$, the first term of Eq.~\eqref{eq:msv_steady_limit} vanishes and the value of steady-state MSV becomes $\frac{2 k_B T}{m}$, resulting in the equipartition theorem. 

Further, we have exactly calculated the steady-state probability distribution function for velocity using the correlation matrix method~\cite{van2007stochastic}. Due to the linearity of the dynamics and the Gaussian nature of the active force, the probability distribution function is Gaussian. Thus, by introducing the correlation matrix $C$ with components
\begin{equation}
    [C_{i,j}] = \langle V_i V_j \rangle - \langle V_i \rangle \langle V_j \rangle,
\end{equation}
one can write the probability distribution function $\mathcal{P}({\bf{V}})$~\cite{van2007stochastic} as
\begin{equation}
\mathcal{P}({\bf{V}}) =  \frac{1}{(2\pi)^4 \sqrt{\text{Det}(C)}} e^{\left[{-\frac{1}{2}\left( \bf{V}^T\right) 
C^{-1}\left(\bf{V} \right)}\right]}.
      \label{eq:total_pdf}
\end{equation}
Here, $\bf{V}$ is the column vector whose elements $V_i \in \left\{v_x, v_y, \xi_x, \xi_y \right\}$ in which the variables  $v_x$, $v_y$, $\xi_x$, and $\xi_y$ are the $x$ and $y$ components of the velocity and the active force, respectively.
The dynamics of the model Eq.~\eqref{eq:model dynamics} and Eq.~\eqref{eq:model dynamics2} can be written in the form
\begin{equation}
    \dot{\bf{V}} = -A \cdot \bf{V}  + B \cdot \boldsymbol{ \eta'}.
    \label{eq:dynamics-matrix}
\end{equation}
The matrices $A$, $B$ and $\boldsymbol{\eta'}$ are given by
\begin{equation}
A=\begin{bmatrix}
 \frac{1}{t_m} & 0 & -\frac{1}{\gamma  t_m} & 0 \\
 0 & \frac{1}{t_m} & 0 & -\frac{1}{\gamma  t_m} \\
 0 & 0 & \frac{1}{t_c} & -\omega  \\
 0 & 0 & \omega  & \frac{1}{t_c} \\
\end{bmatrix} ,
\end{equation}
\begin{equation}
  B=\begin{bmatrix}
 \frac{1} {t_m}\sqrt{\frac{2 k_BT}{\gamma }}& 0 & 0 & 0 \\
 0 & \frac{1} {t_m}\sqrt{\frac{2 k_BT}{\gamma }} & 0 & 0 \\
 0 & 0 & \xi _0 \sqrt{\frac{2}{t_c}} & 0 \\
 0 & 0 & 0 & \xi _0 \sqrt{\frac{2}{t_c}} \\
\end{bmatrix} ,
\end{equation}
and 
\begin{equation}
    \boldsymbol{\eta'} = (\eta_x\ \eta_y\ \zeta_x\ \zeta_y)^T.
\end{equation}
Following the correlation matrix formalism, the correlation matrix $C$ can be shown to satisfy the equation
\begin{equation}
    A\cdot C + C \cdot A^T + B\ B^T = 0.
    \label{eq:C_relation}
\end{equation}
\begin{figure*}
    \includegraphics[width=0.8\linewidth]{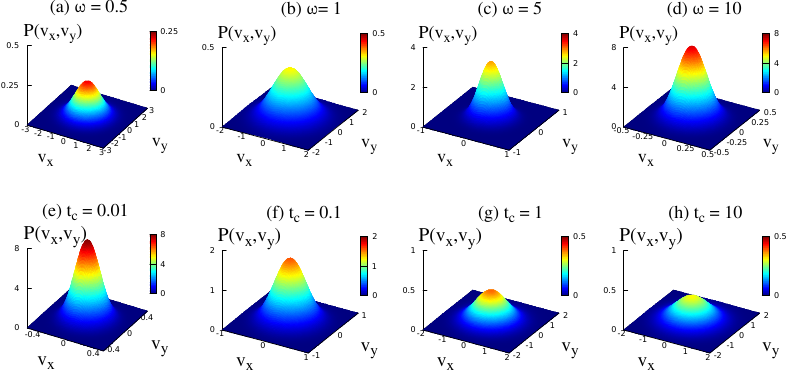}
    \caption{The steady state  probability distribution  of velocity [Eq.~\eqref{eq:pdf_Pv}] for different values of $\omega$ fixing $t_c(=5.0)$ are shown in (a)-(d) and for different values of $t_c$ for a fixed $\omega=1.0$ is shown in (e)-(h). Other parameters are $t_m = \gamma = \xi_0 = 1$ and $T = 0.01$.}
    \label{fig:PD_v}
\end{figure*}
By substituting the solution of C from Eq.~\eqref{eq:dynamics-matrix} into Eq.~\eqref{eq:total_pdf}, and integrating over all other variables, the steady state probability distribution function for the velocity $P(v_x,v_y)$ can be obtained as
\begin{equation}
    P(v_x, v_y)=\frac{1}{2\pi\sigma^2} \exp{\left(-\frac{ v_x^2 + v_y^2}{2\sigma^2}\right)},
    \label{eq:pdf_Pv}
\end{equation}
with the variance $\sigma^2$ given by
\begin{equation}
   \sigma^2 = \frac{\xi _0^2 t_c t_m \left(t_c+t_m\right)+ \gamma  k_BT \left(\omega ^2 t_c^2 t_m^2+\left(t_c+t_m\right){}^2\right)}{\gamma ^2 t_m \left(\omega ^2 t_c^2 t_m^2+\left(t_c+t_m\right){}^2\right)}.
\end{equation}
Figure~\ref{fig:PD_v} shows the steady-state probability distribution of the velocity for different values of $\omega$ and $t_c$. The distribution is Gaussian centered at the origin.
As $\omega$ increases, $\sigma^2$ decreases. Hence, the width of the distribution decreases and the distribution narrows out [see Figs.~\ref{fig:PD_v} (a)-(d)]. 
This result agrees with the steady-state MSV calculation [Eq.~\eqref{eq:msv_steady_limit}]. This is because, as $\omega$ increases, the number of circles appearing in the trajectory of the particle increases, which shrinks the overall displacement and hence results in the decrease of MSV, confirming the transient confinement of the particle. 
However, as the value of $t_c$ increases, the probability distribution of velocity spreads [see Figs.~\ref{fig:PD_v} (e)-(h)].

\begin{figure}
    \centering
    \includegraphics[width=\linewidth]{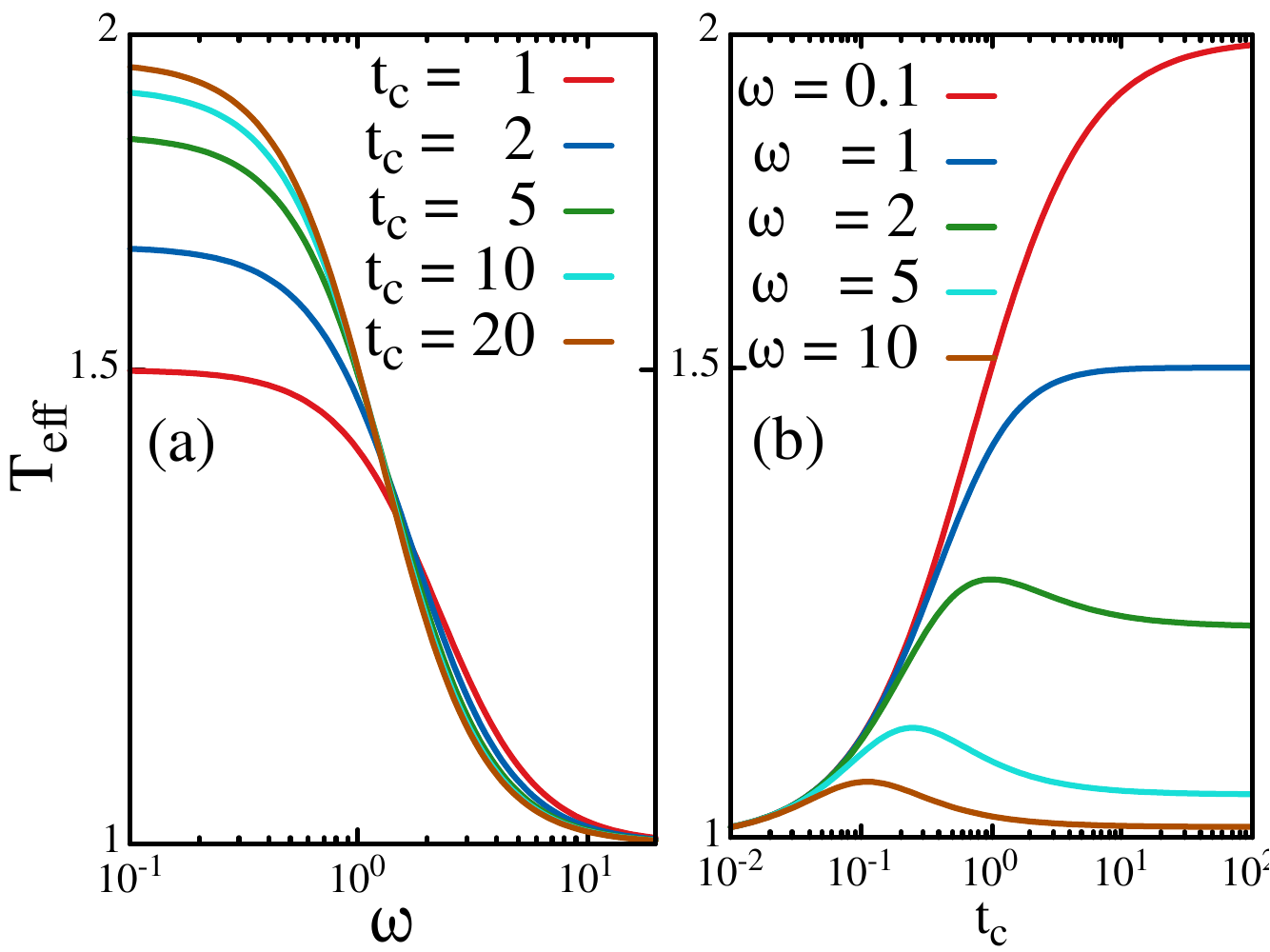}
    \caption{The effective temperature $\Teff$[Eq.~\eqref{eq:T_eff}] is plotted in (a) as a function of $\omega$ for different values of $t_c$, and in (b) as a function of $t_c$ for different values of $\omega$. Other common parameters are $ T =t_m = \gamma = \xi_0 = 1$.}
    \label{fig:T_eff}
\end{figure}

The steady state MSV $\langle \Delta{\bf v}^2 \rangle_s$ can also be calculated using the correlation matrix $C$ as 
\begin{equation}
    \begin{split}
      \langle \Delta{\bf v}^2 \rangle_s &= C_{1,1} + C_{2,2}\\
     & = \frac{2} {\gamma ^2}\left(\frac{\xi _0^2 t_c \left(t_c+t_m\right)}{t_c^2 \left(\omega ^2 t_m^2+1\right)+2 t_c t_m+t_m^2}+\frac{\gamma k_B T}{t_m}\right).
    \end{split}
    \label{eq:msv_st_eq}
\end{equation}
This is consistent with Eq.~\eqref{eq:msv_steady_limit}.
From the analysis of the MSD, MSV, position and velocity distribution functions, and the particle trajectories, it is clear that there exists a chirality-induced dynamic confinement for the self-propulsion of the particle. The effect of this chirality induced confinement on the dynamics of the particle can be understood by introducing the concept of effective kinetic temperature~\cite{loi2008effective, caprini2024self, arsha2024inertial, adersh2024inertial}.
The stationarity of the stochastic process ${\bf v}(t)$ allows us to define the effective temperature ($\Teff$) of the system~\cite{sevilla2018nonequilibrium} as
\begin{equation}
    k_B \Teff = \frac{1}{2} m \langle \Delta {\bf v}^2 \rangle_s.
    \label{eq:Teff_def}
\end{equation}
Substituting $\langle \Delta{\bf v}^2 \rangle_s$ from Eq.~\eqref{eq:msv_st_eq} in Eq.~\eqref{eq:Teff_def}, we get
\begin{equation}
     \Teff = T + \frac{t_m} {k_B\gamma  }\left(\frac{\xi _0^2 t_c \left(t_c+t_m\right)}{t_c^2 \left(\omega ^2 t_m^2+1\right)+2 t_c t_m+t_m^2}\right).
    \label{eq:T_eff}
\end{equation}
Thus, the $\Teff$ has two parts. The first part is simply the temperature $T$ of the viscous bath. The second part of Eq.~\eqref{eq:T_eff} is basically due to the active transport and it depends both on $t_c$ and $\omega$.
In the equilibrium limit, that is, in either of the limits $t_c \to 0$ or $\omega \to \infty$, we get
\begin{equation}
     \lim_{t_c \to 0}\Teff = \lim_{\omega \to \infty}\Teff = T.
\end{equation}
The variation of $\Teff$ as a function of $\omega$ and $t_c$ is shown in Figs.~\ref{fig:T_eff}(a) and~\ref{fig:T_eff}(b),  respectively. In $\omega \to 0$ limit, $\Teff$ is given by
\begin{equation}
    \lim_{\omega \to 0}\Teff = T + \frac{\xi_0^2}{k_B\gamma} \left( \frac{t_c t_m}{t_c + t_m}\right).
\end{equation}
This constant value of $\Teff$, which corresponds to the effective temperature of an achiral active particle, is maintained for lower values of $\omega$ and for a finite $t_{c}$. Notably, $\Teff$ increases from the equilibrium temperature $T$ in the limit of small $t_c$ to the value $T + \frac{\xi_0^2 t_m}{k_B \gamma}$ as $t_c \to \infty$, reflecting the influence of inertial effects on the system.
With further increase in $\omega$ value, $\Teff$ decays to the equilibrium temperature $T$ [Fig.~\ref{fig:T_eff}(a)]. This decrease of $\Teff$ as a function of chirality is also observed in Ref.~\onlinecite{caprini2023chiral}. 
This is because, for higher $\omega$ values or in the $\omega \rightarrow \infty$ limit, the second term of Eq.~\eqref{eq:T_eff} vanishes, and $\Teff$ approaches the equilibrium temperature of the bath. 
Similarly, by increasing $t_c$ values, for lower $\omega$ values, $\Teff$ initially increases with an increase in $t_{c}$ and saturates for larger $t_{c}$ values. However, for larger $\omega$, $\Teff$ shows a non-monotonic dependence of $t_{c}$. It increases, shows a maximum, and then decreases with $t_{c}$ and saturates to a constant value [see Fig.~\ref{fig:T_eff}(b)].
This value can be obtained from Eq.~\eqref{eq:T_eff} by taking the limit $t_c \to \infty$, and we obtain
\begin{equation}
    \lim_{t_c \to \infty} \Teff = T + \frac{t_m \xi_0^2}{k_B \gamma (1 + t_m^2\omega^2)}.
\end{equation}
This decline in $\Teff$ with increasing $\omega$ can be attributed to the transient confining mechanism of the particle, as previously discussed. Finally, for very large $\omega$ values (i.e., $\omega \rightarrow \infty$ limit), $\Teff$ approaches to the equilibrium temperature $T$.

\section{SUMMARY}\label{sec:summary}

In summary, we have explored the self-propulsion of an inertial chiral active Ornstein-Uhlenbeck particle in two dimensions. Chirality is incorporated into the system by modulating the Ornstein-Uhlenbeck process and introducing a coupling term. 
This straight forward model of a chiral particle preserves the linearity of the AOUP framework, facilitating more effective analytical treatment. 
Using both exact analytical calculations and numerical simulations, we examined the transient and steady-state transport properties of the particle. 
From the simulated particle trajectories, it is revealed that for a finite chirality and duration of activity, there is an emergence of circular motion in the particle trajectory. 
Notably, the radius of circular motion decreases as the magnitude of chirality increases. As a result, the transient trajectory of the particle becomes confined to a smaller region for higher magnitude of chirality. Additionally, the increase in the duration of activity also reduces the overall displacement of the particle.

Further, the exact calculation of MSD shows an initial time ballistic and long time or steady-state diffusive behavior. 
The initial ballistic regime is independent of the magnitude of chirality and activity timescale, where as the steady-state MSD decreases with either increase in magnitude of chirality or duration of activity. In addition to the typical ballistic and diffusive behaviors, unlike an active Brownian particle, a chiral particle interestingly manifests an intermediate time sub-diffusive behavior at lower chirality regime and it transitions to a non-diffusive behavior for higher chirality. Moreover, this intermediate non-diffusive region is more pronounced in the over-damped limit, where inertial impact is negligible.
This observation confirms the presence of a chirality-induced transient confinement of the particle that persists for a short interval of time and diffuses away in the time asymptotic limit or steady state. These results are further complemented by the position probability distribution, which gets narrower and sharper as either chirality or duration of activity is increased. Moreover, the emergence of oscillations in the intermediate regime of the MSD might be attributed due to the intricate interplay between the activity and chirality. 

The characteristic feature of this chirality-induced transient confinement of the particle is further corroborated by the precise calculation of both mean square velocity and effective kinetic temperature of the medium, which decrease as a function of the magnitude of chirality. Finally, we believe that our study provides a detailed understanding of how chirality influences the properties of active particles, offering valuable insights into the dynamics and steady-state behavior of various natural and artificial chiral active systems.

\section{Acknowledgement}
 MS acknowledges the start-up grant from UGC, SERB-SURE grant (SUR/2022/000377), and CRG grant (CRG/2023/002026) from DST, Govt. of India for financial support.

\end{document}